\documentclass{article}
\usepackage[pdftex]{graphicx}

\usepackage[preprint]{neurips_2025}

\usepackage[utf8]{inputenc} 
\usepackage[T1]{fontenc}    
\usepackage{hyperref}       
\usepackage{url}            
\usepackage{booktabs}       
\usepackage{amsfonts}       
\usepackage{nicefrac}       
\usepackage{microtype}      
\usepackage{xcolor}         

\title{Boltzina: Efficient and Accurate Virtual Screening via Docking-Guided Binding Prediction with Boltz-2}

\author{%
  Kairi Furui \\
  Department of Computer Science\\
  School of Computing\\
  Institute of Science Tokyo\\
  Kanagawa, 226-8501, Japan \\
  \texttt{furui@li.comp.isct.ac.jp} \\
  \And
  Masahito Ohue \\
  Department of Computer Science\\
  School of Computing\\
  Institute of Science Tokyo\\
  Kanagawa, 226-8501, Japan \\
  \texttt{ohue@comp.isct.ac.jp} \\
}

\begin{document}

\maketitle

\begin{abstract}
  In structure-based drug discovery, virtual screening using conventional molecular docking methods can be performed rapidly but suffers from limitations in prediction accuracy. Recently, Boltz-2 was proposed, achieving extremely high accuracy in binding affinity prediction, but requiring approximately 20 seconds per compound per GPU, making it difficult to apply to large-scale screening of hundreds of thousands to millions of compounds.
  This study proposes Boltzina, a novel framework that leverages Boltz-2's high accuracy while significantly improving computational efficiency.
  Boltzina achieves both accuracy and speed by omitting the rate-limiting structure prediction from Boltz-2's architecture and directly predicting affinity from AutoDock Vina docking poses.
  We evaluate on eight assays from the MF-PCBA dataset and show that while Boltzina performs below Boltz-2, it provides significantly higher screening performance compared to AutoDock Vina and GNINA.
  Additionally, Boltzina achieved up to 11.8$\times$ faster through reduced recycling iterations and batch processing.
  Furthermore, we investigated multi-pose selection strategies and two-stage screening combining Boltzina and Boltz-2, presenting optimization methods for accuracy and efficiency according to application requirements.
  This study represents the first attempt to apply Boltz-2's high-accuracy predictions to practical-scale screening, offering a pipeline that combines both accuracy and efficiency in computational biology. 
  The Boltzina is available on github; \url{https://github.com/ohuelab/boltzina}.
\end{abstract}

\section{Introduction}
In drug discovery research, structure-based drug design (SBDD) is a method that utilizes three-dimensional structural information of target proteins to design and evaluate novel compounds~\cite{Saini2025-um}.
Among these approaches, virtual screening (VS)~\cite{Lionta2014-qh} has been widely used to select promising candidate molecules from vast compound libraries.
In conventional VS, pose generation by molecular docking and binding affinity prediction using scoring functions have been standard methods~\cite{Morris2009-cq,Trott2010-aq}, but scoring functions based on physical models or empirical rules have limitations in accuracy~\cite{Valsson2025-zk,Shirali2025-pg}.

To address this challenge, machine learning-based scoring functions (MLSFs) have been proposed~\cite{McNutt2021-nq,Li2021-wg,Durant2025-jq}.
Diverse MLSF approaches range from classical Random Forests using interaction features~\cite{Li2015-wz} to neural network methods such as convolutional networks based on contact information~\cite{Wang2021-tn} and graph neural networks operating on atomic and interaction graphs~\cite{Jiang2021-tp}.
While MLSFs sometimes show higher accuracy than conventional methods, they still face issues such as data dependency and insufficient generalization to unknown targets, leaving reliability challenges in actual drug discovery applications~\cite{Durant2025-jq,Zhu2022-xi}.

The recently proposed Boltz-2~\cite{Passaro2025-tw} integrates structure prediction and binding affinity prediction using an AlphaFold3-like~\cite{Abramson2024-mf} diffusion model, achieving performance that significantly exceeds conventional docking-based methods and MLSFs.
Boltz-2 demonstrates performance approaching molecular simulation-based free energy calculations~\cite{Wu2025-ru,Ross2023-qj} by incorporating an affinity module for protein--ligand complexes in addition to Boltz-1's~\cite{Wohlwend2024-oq} structure prediction capabilities.
Furthermore, evaluation using MF-PCBA~\cite{Buterez2023-ux} reported screening performance that significantly surpasses existing compound-protein interaction (CPI) prediction models~\cite{Li2022-ag} and empirical scoring methods~\cite{McGann2011-dy}.

However, Boltz-2 requires approximately 20 seconds per ligand for prediction, making direct application to large-scale libraries exceeding one million compounds impractical~\cite{Yu2023-wz,Lyu2023-qt}.
This is because Boltz-2 requires a diffusion process for structure prediction, which causes the computational time bottleneck in affinity prediction.
To solve this challenge, methods that maintain Boltz-2's high accuracy while improving computational efficiency are required.

This study proposes Boltzina, a framework that achieves rapid compound screening by directly predicting Boltz-2's affinity and binding from poses generated by the existing docking method AutoDock Vina~\cite{Trott2010-aq}.
We compare these methods on the MF-PCBA dataset and examine their applicability to large-scale screening.
Furthermore, we decompose Boltzina's components to clarify the accuracy–speed trade-off, and evaluate multi-pose selection strategies and two-stage screening with Boltz-2 to discuss applicability in actual large-scale screening.

\section{Materials and Methods}
\subsection{Boltzina Pipeline}
\begin{figure}[tbp]
  \centering
  \includegraphics[width=\textwidth]{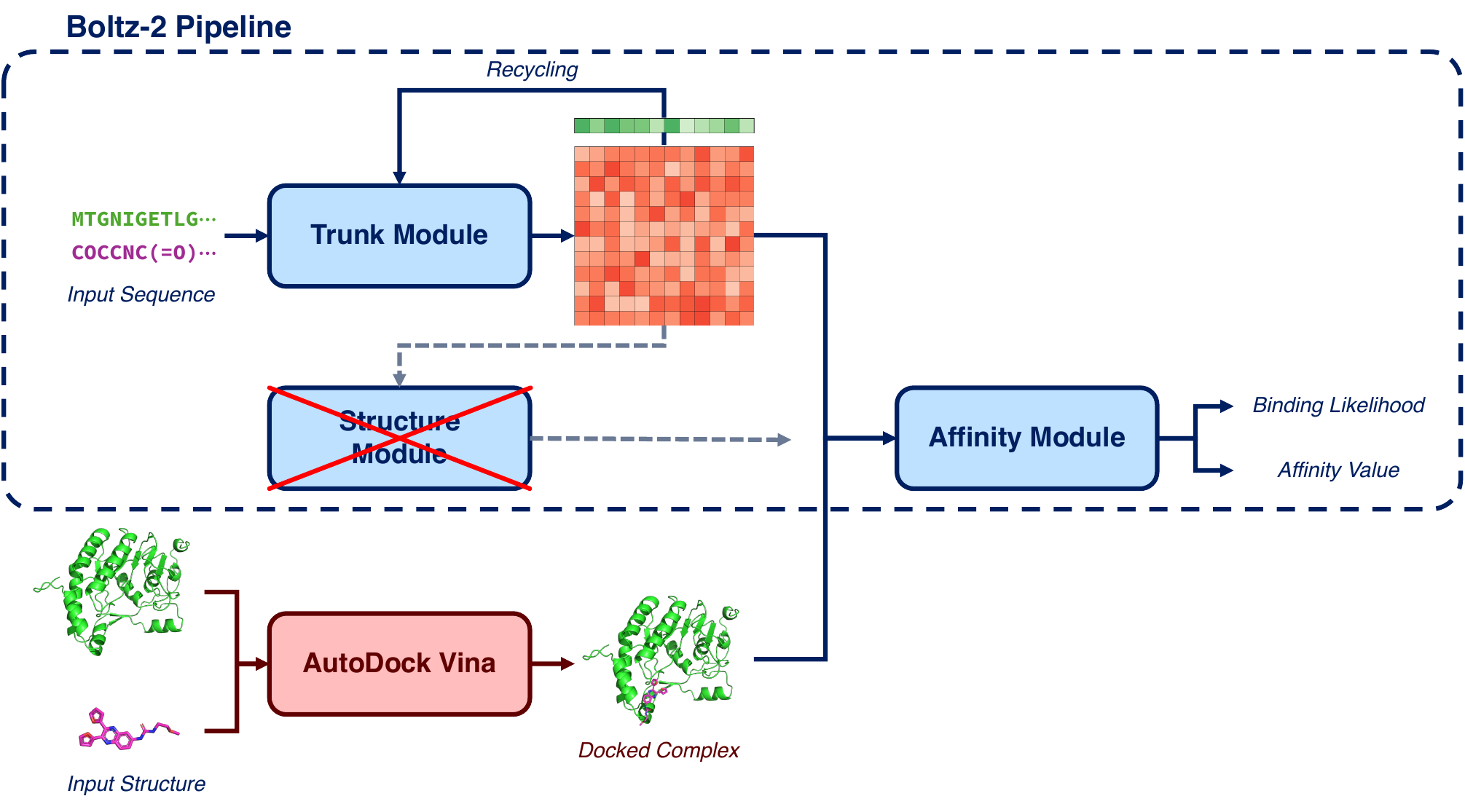}
  \caption{Overview of the Boltzina pipeline. The dashed blue box represents the original Boltz-2 affinity prediction pipeline, and the red $\times$ mark on the structure module indicates that it is omitted in Boltzina.}
  \label{fig:boltzina-pipeline}
\end{figure}

Boltzina is based on Boltz-2's architecture as shown in Figure~\ref{fig:boltzina-pipeline}.
In the original Boltz-2, binding affinity is predicted through staged information processing using the trunk module, structure module, and affinity module.
First, the trunk module extracts latent structural features from input protein sequences and ligands.
The trunk module primarily consists of PairFormer and MSA modules, generating pairwise representations that capture intermolecular interactions.
This latent representation implicitly contains structural information.
Next, the structure module predicts 3D structures based on the latent representation from the trunk module, determining atomic coordinates and the geometric arrangement of protein--ligand complexes.
Finally, the affinity module predicts binding affinity using intermolecular interaction information obtained from the trunk module and explicit 3D coordinate information predicted by the structure module.
The affinity module performs processing specialized for protein--ligand interface interactions through a dedicated PairFormer architecture, outputting two predicted values: binding likelihood and affinity value.
In Boltzina, instead of using the 3D coordinate information from the structure module, poses generated by rigid docking with the external docking software AutoDock Vina are directly input to the affinity module, thereby omitting the structure module.
This allows Boltzina to retain Boltz-2's high-accuracy intermolecular interaction analysis capabilities while avoiding the computational cost required for structure prediction.

In our implementation, PDB files of docking structures generated by AutoDock Vina were converted to appropriate MMCIF-format complex structures using PDB-tools~\cite{Rodrigues2018-hr} and the MAXIT suite~\cite{MAXIT_RCSB}.
These converted structures were then processed into the model's input format by reusing Boltz-2's template structure processing implementation.
Furthermore, we constructed a batch processing pipeline for efficiently handling multiple complexes and improved processing efficiency by making the batch size variable, whereas it was fixed at 1 in Boltz-2.

\subsection{Dataset}
For evaluation, we used a test set independently constructed from the MF-PCBA dataset~\cite{Buterez2023-ux}, following~\cite{Passaro2025-tw}.
This dataset is a virtual screening benchmark for developing and evaluating machine learning methods in drug discovery, containing multiple targets collected from PubChem, and was also used in Boltz-2 to evaluate model screening performance.
In this study, we conducted experiments on 8 out of 10 MF-PCBA assays evaluated in Boltz-2, as listed in Table~\ref{tab:datasets}.
Protein sequences similar to these test data were appropriately filtered in Boltz-2's training for affinity prediction tasks~\cite{Passaro2025-tw}.
Here, AID489030 was excluded because the clear binding pocket was unknown and the grid could not be determined, and AID485273 was excluded because the active ligands included a high proportion of large molecules with more than 60 heavy atoms, which would compromise pose estimation by AutoDock Vina.
In preprocessing, after applying PAINS (Pan-Assay Interference Compounds) filtering~\cite{Baell2010-gd}, all binders and non-binders were randomly sampled to total 50{,}000 cases, then duplicates were removed.
Furthermore, molecules for which AutoDock Vina failed (e.g., those containing arsenic atoms) were excluded, and screening performance was finally evaluated with the numbers shown in Table~\ref{tab:datasets}.
To evaluate execution time and the effects of using multiple docking poses, 1{,}000 ligands were randomly sampled from the MF-PCBA test set.

\begin{table}[tbp]
  \centering
  \caption{Information on the 8 MF-PCBA test set assays used for evaluation and their corresponding ligands. PDB IDs were used as references for grid positions when available. Active and Inactive indicate the number of active and inactive compounds in each assay.}
  \label{tab:datasets}
  \begin{tabular}{cccc}
    \toprule
    PubChem AID & PDB ID/CID & Active & Inactive \\
    \midrule
    743445 & 6UE6 & 144 & 49851 \\
    485317 & 3MBG & 976 & 49004 \\
    2097 & 1J1B & 522 & 49393 \\
    493091 & 3PGL & 782 & 49203 \\
    2650 & 7SXF & 612 & 49270 \\
    504329 & CID1474141 & 466 & 49516 \\
    588689 & 3EVG & 486 & 49499 \\
    588549 & CID2277079 & 159 & 49828 \\
    \bottomrule
  \end{tabular}
\end{table}

\subsection{Docking Settings}
AutoDock Vina~\cite{Eberhardt2021-jr} v1.2.7 was used for docking pose generation.
The docking grid size was set to 20~\AA{} with exhaustiveness set to 8, and for the main screening performance evaluation, only the best pose predicted by AutoDock Vina was evaluated with Boltz-2.
Docking was performed using one protein--ligand complex structure predicted by Boltz-2 as a reference structure for each target.
For the six assays where clear known complex structures existed in the PDB database, those ligands were used; for the remaining two assays, the most potent ligand among MF-PCBA binders was used as input to predict holo structures (Table~\ref{tab:datasets}).
Because Vina requires explicit specification of binding pockets, the centroid of the holo ligand was used as the grid center.
We confirmed that most binders in Boltz-2's complex prediction structures bound to the same region as the aforementioned predicted holo structures, validating the appropriateness of the grid position.

The execution environment used a machine with one H100 GPU and 48 CPU cores.
Docking calculations were executed in 48 parallel processes, assuming actual screening scenarios.
Execution time measurements were performed under these parallelization conditions.

\subsection{Comparison Methods}
For screening performance evaluation, we compared Boltzina's performance with the following methods:
\begin{description}
  \item[Boltz-2] Predictions using the original Boltz-2 with default settings.
  \item[AutoDock Vina] Conventional molecular docking using the above settings.
  \item[GNINA] Open-source software GNINA v1.3.2~\cite{McNutt2021-nq,McNutt2025-dw}, which incorporates CNN-based scoring functions on top of AutoDock Vina and Smina~\cite{Koes2013-hb}. GNINA ranked nine poses generated by AutoDock Vina using CNN VSScore (the product of CNN affinity and CNN score)~\cite{Dunn2024-ve}, and selected the highest-scoring pose.
\end{description}

We also evaluated two parameters to examine each component's contribution to performance and effects on execution speed:
\begin{description}
  \item[Boltzina (Cycle=1)] Recycling in the trunk module reduced from the default five iterations to one (see Figure~\ref{fig:boltzina-pipeline}).
  \item[Boltzina (No Pose)] Docking pose information masked by setting all ligand atom coordinates to the origin, eliminating initial pose dependency. Since docking pose information is not used, docking time is excluded from execution time.
\end{description}

\subsection{Pose Selection Strategies}
To evaluate the impact of using multiple docking poses, we tested the following strategies:
\begin{description}
  \item[Best Pose Only] Evaluating only the best (minimum-energy) pose from AutoDock Vina.
  \item[Top-$N$ Best Score] Selecting the highest Boltzina binding likelihood among the top $N$ poses generated by AutoDock Vina.
  \item[Top-$N$ Average] Ranking by the average of the Boltzina predicted affinity scores over the top $N$ poses.
\end{description}
Pose selection strategies were evaluated by randomly sampling 1{,}000 molecules per assay.

\subsection{Two-Stage Screening Experiment}
To achieve an optimal balance between computational efficiency and prediction accuracy, we considered a hierarchical screening strategy~\cite{kumar2015hierarchical} combining Boltzina and Boltz-2.
In this strategy, methods with different computational costs are combined in stages.
In the first stage, all compounds are rapidly screened using Boltzina; in the second stage, only promising compounds ranked highly are re-evaluated in detail with the more accurate Boltz-2.
Specifically, the top $p\%$ of compounds were selected based on Boltzina's binding likelihood, these were rescored with Boltz-2, and final rankings were determined.
Evaluation was performed under four conditions with $p$ values of 50\%, 20\%, 10\%, and 5\%.

\section{Results and Discussion}
\subsection{Comparison with Existing Methods}
\begin{figure}[tbp]
  \centering
  \includegraphics[width=\textwidth]{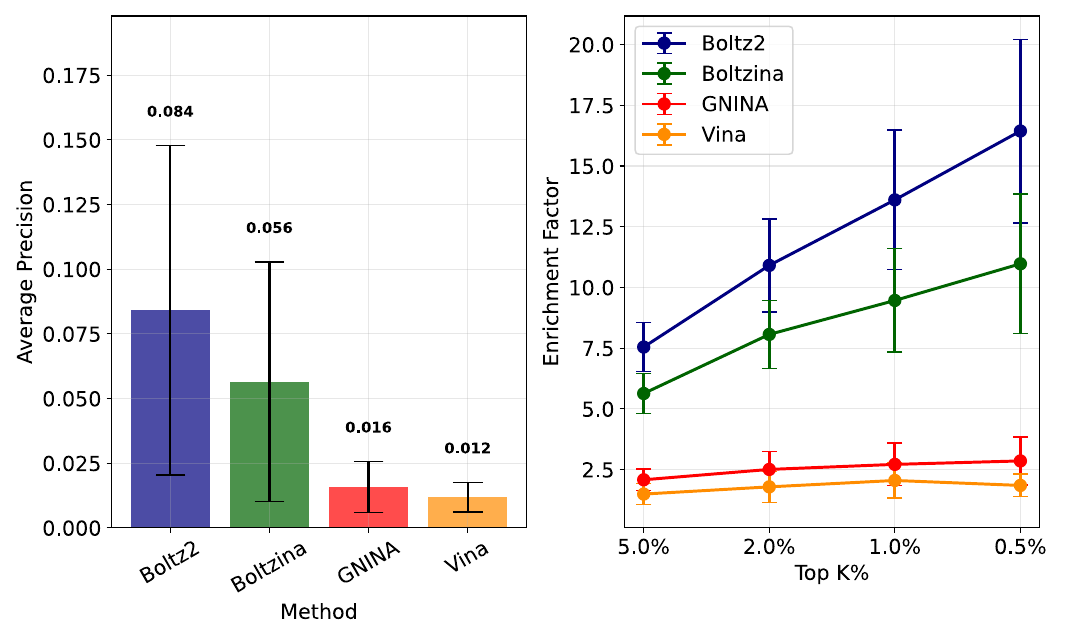}
  \caption{Comparison with existing methods on the MF-PCBA test set. a) mean Average Precision across assays. b) Average Enrichment Factor at top $K\%$ ($K = 0.5\%, 1\%, 2\%, 5\%$).}
  \label{fig:AP_EF}
\end{figure}

We evaluated the proposed method's screening performance using eight assays from the MF-PCBA dataset.
As shown in Figure~\ref{fig:AP_EF}a, for Average Precision (AP), Boltz-2 showed the highest performance (mean AP~0.084), followed by Boltzina (mean AP~0.056).
In contrast, the mean AP of existing GNINA and Vina methods was extremely low.
Therefore, Boltzina achieved significant performance improvement compared to AutoDock Vina and GNINA.

\begin{figure}[tbp]
  \centering
  \includegraphics[width=\textwidth]{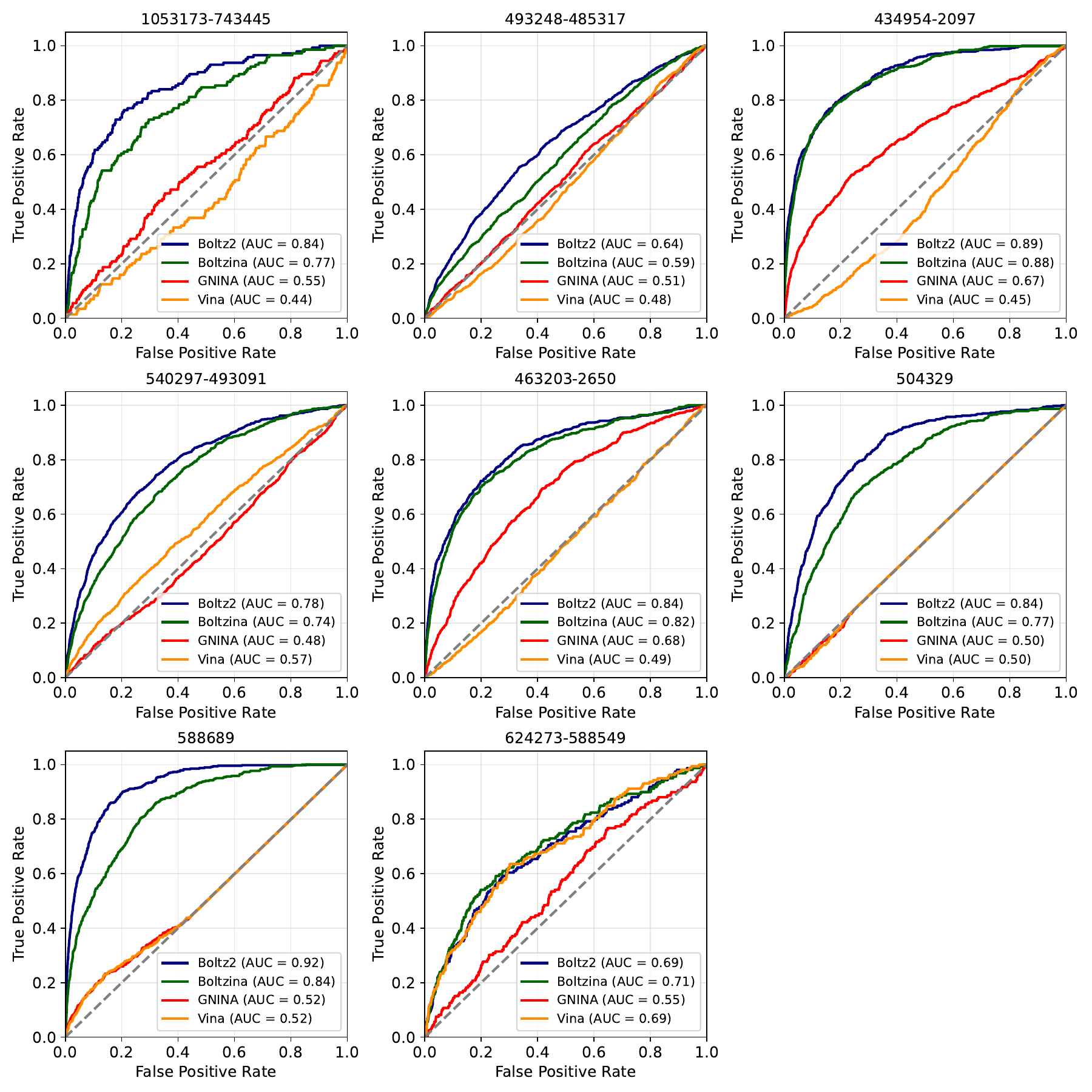}
  \caption{ROC curves for each assay in the MF-PCBA test set.}
  \label{fig:roc_auc_target}
\end{figure}

The Enrichment Factor (EF) results shown in Figure~\ref{fig:AP_EF}b also exhibited similar trends.
While GNINA showed slight improvement over AutoDock Vina through reranking, it was inferior to Boltzina's performance increase.
Additionally, at the top 5\%, the performance difference between Boltz-2 and Boltzina was relatively small, whereas at the top 0.5\% there was a substantial difference between the two methods.
This suggests that Boltz-2's precise calculations are essential for accurate ranking of a very small number of top compounds, while Boltzina is particularly effective for screening scenarios requiring medium-scale enrichment.
Figure~\ref{fig:roc_auc_target} shows ROC curves for each assay.
While AutoDock Vina showed near-random performance for many targets, it improved significantly with Boltz-2 rescoring.
Additionally, for some assays, the performance difference between Boltz-2 and Boltzina was small, indicating cases where Boltzina approached Boltz-2's performance.

\begin{figure}[tbp]
  \centering
  \includegraphics[width=\textwidth]{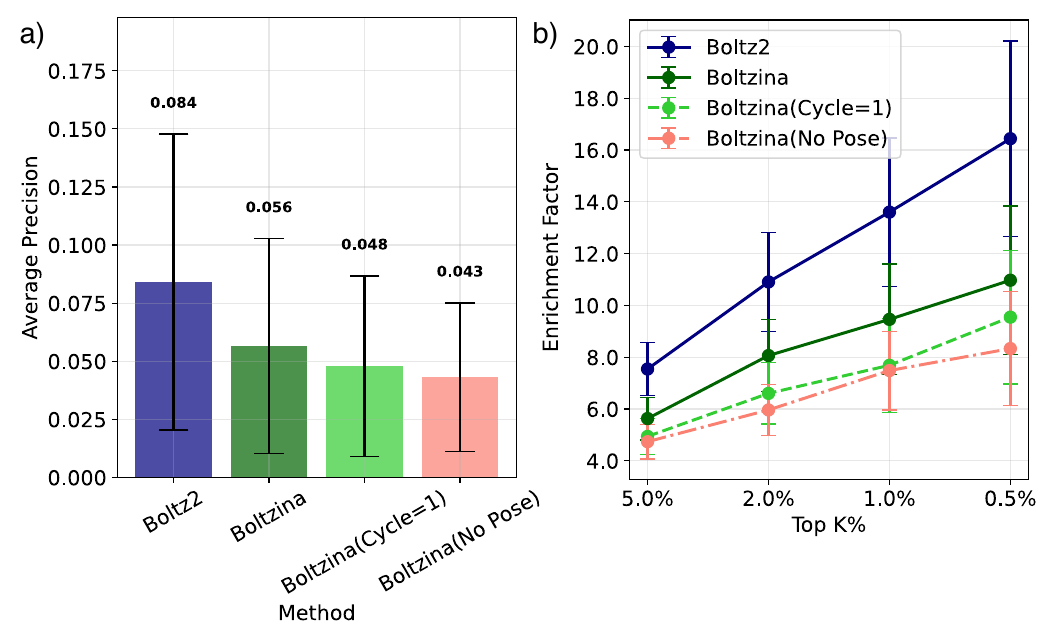}
  \caption{Comparison of Boltzina variants on the MF-PCBA test set. a) Mean Average Precision across assays. b) Average Enrichment Factor at top $K\%$.}
  \label{fig:AP_EF_boltzina}
\end{figure}

Next, Figure~\ref{fig:AP_EF_boltzina} shows comparisons of different parameters for Boltzina.
For Boltzina (Cycle=1), the mean AP decreased from 0.056 to 0.048, but the decrease was limited, confirming that reducing the number of recycling iterations is effective for cutting computational cost without significantly compromising accuracy.
On the other hand, for Boltzina (No Pose), even when ligand pose information was not provided, the mean AP was 0.043; while there was a negative impact on performance, it was more limited than expected.
This result suggests that intermolecular interaction information obtained from the trunk module plays an important role in binding prediction.
That is, the reason performance improves without depending on AutoDock Vina's pose accuracy is suggested to be the latent representations of intermolecular interactions learned by the trunk module.

Next, as shown in Figure~\ref{fig:exec_time_rocauc}a, Boltzina achieved a significant reduction in execution time.
The average processing time per ligand was approximately 16.5~seconds for Boltz-2, compared to 2.3~seconds for Boltzina, which is 7.3$\times$ faster, and 1.4~seconds for Boltzina (Cycle=1), which is 11.8$\times$ faster.
The main factors for this speedup are omission of structure prediction steps, improved parallel processing efficiency through increased batch size, and reduced recycling iterations.
Notably, AutoDock Vina docking required approximately 0.8~seconds and became rate-limiting for Boltzina (Cycle=1).
Therefore, the main bottlenecks for Boltzina are docking pose generation and trunk module processing.
By adjusting these parameters according to compound library scale and usage, the balance between accuracy and speed could be further optimized.
Overall, Boltzina combines practical computational efficiency with screening performance, making it a promising method especially for large-scale screening where time efficiency matters more than the need for high accuracy.

\subsection{Pose Selection Strategies}
\begin{figure}[tbp]
  \centering
  \includegraphics[width=\textwidth]{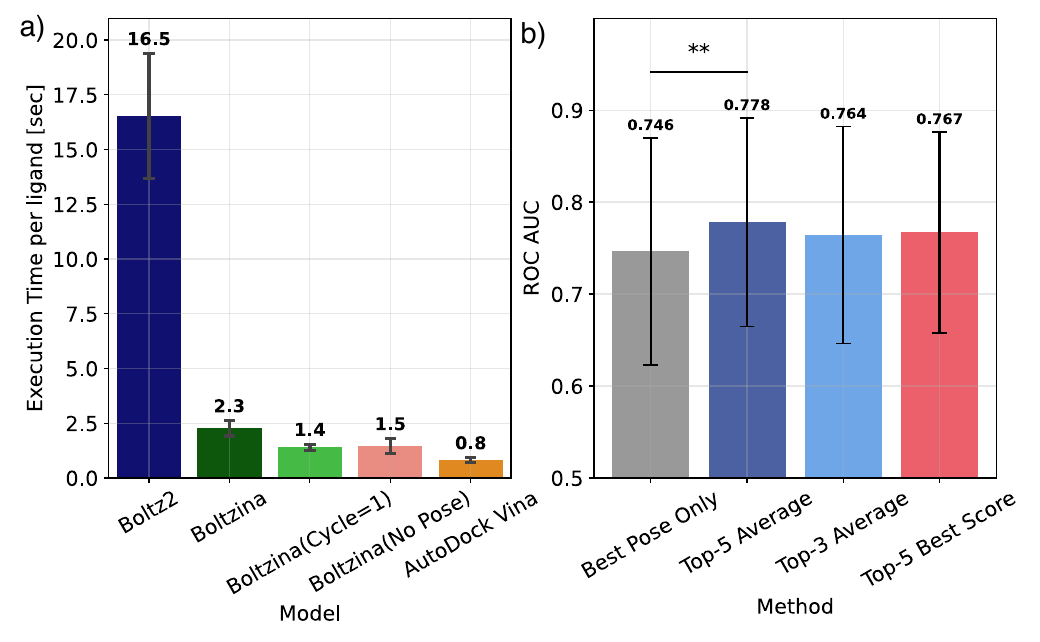}
  \caption{a) Execution time per ligand, computed from 1{,}000 ligands in the MF-PCBA test set using one GPU and 48 CPU cores. b) Average ROC-AUC for pose selection strategies using 1{,}000 ligands from the MF-PCBA test set. ${**}$ indicates a significant difference ($p<0.01$) by the Wilcoxon signed-rank test.}
  \label{fig:exec_time_rocauc}
\end{figure}

Figure~\ref{fig:exec_time_rocauc}b shows the average ROC-AUC for each pose selection strategy.
Top-5 Average strategy showed the highest ROC-AUC (0.778), improving performance compared to Top-3 Average and Best Pose Only strategies (0.746).
Additionally, Top-5 Best Score strategy did not outperform the averaging strategy.
If Boltz-2's binding likelihood could accurately identify the quality of individual poses, selecting the highest score should be superior; in practice, averaging was advantageous.
This suggests that Boltz-2's binding likelihood may not sufficiently discriminate individual pose quality, and the improvement from averaging likely stems from reduced randomness.
Such behavior is likely attributed to the fact that Boltz-2's binding likelihood is not trained to discriminate the quality of individual poses.

\subsection{Two-Stage Screening Experiment}
\begin{figure}[tbp]
  \centering
  \includegraphics[width=.85\textwidth]{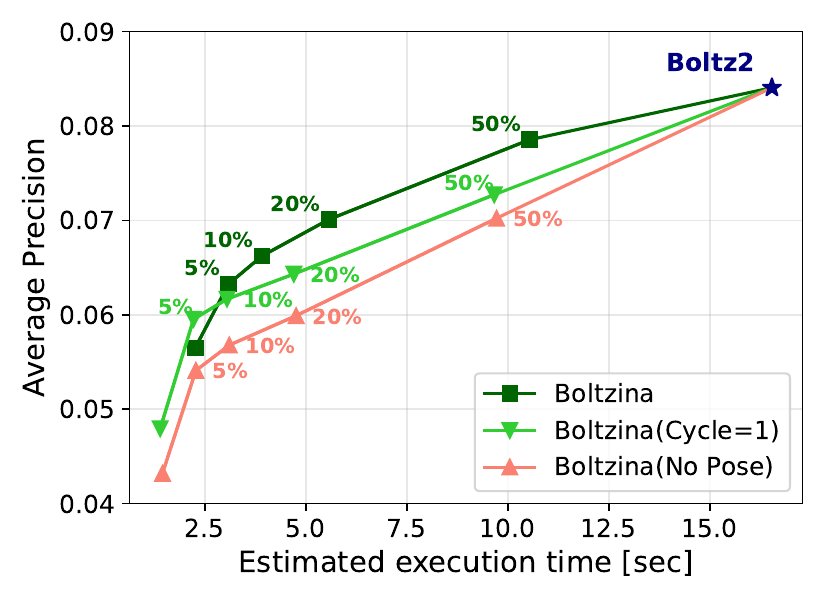}
  \caption{Two-stage screening: estimated execution time vs. mean Average Precision at different selection ratios. Estimated time was computed from the per-ligand measurements in Figure~\ref{fig:exec_time_rocauc}a.}
  \label{fig:two_step_pareto}
\end{figure}

Finally, we conducted two-stage screening experiments combining rapid pre-screening with Boltzina and accurate binding prediction with Boltz-2.
Figure~\ref{fig:two_step_pareto} shows the relationship between estimated execution time and Average Precision for each method.
Comparing Boltzina variants, Boltzina provided Pareto-optimal solutions in most cases.
In particular, combining Boltzina (Cycle=1) with rescoring the top 5\% by Boltz-2 outperformed Boltzina alone in accuracy per unit cost.
Moreover, using the top $\sim$20\% from Boltzina for two-stage screening achieved a mean AP exceeding 0.07 while being about three times faster than Boltz-2.
These results show that two-stage strategies enable finer control of the trade-off between Boltzina and Boltz-2 according to application requirements.
However, the optimal ratio depends on the prevalence of potential binders in the target library and thus should be chosen accordingly in practice.

\section{Conclusion}
We developed Boltzina, a pipeline for rapid and accurate virtual screening that predicts affinity using Boltz-2 with AutoDock Vina docking poses as input.
As a result, while Boltzina showed somewhat lower accuracy than Boltz-2, it achieved substantially better screening performance than conventional molecular docking methods.
Execution time achieved a 7.3$\times$ speedup over Boltz-2 in the standard setting, and up to 11.8$\times$ when recycling was reduced to a single iteration, enabling screening within realistic timeframes for library sizes that are difficult for Boltz-2 alone.

Experimental results indicate that intermolecular interaction representations generated by the trunk module largely govern binding prediction accuracy, allowing reasonable performance even without pose information; providing appropriate poses from AutoDock Vina further improves accuracy.
Averaging multiple poses improved ROC-AUC and effectively reduced randomness associated with pose selection.
Furthermore, two-stage screening that combines rapid screening with Boltzina and precise re-evaluation with Boltz-2 yielded Pareto-optimal solutions in both computational efficiency and accuracy.

Several limitations remain.
We did not evaluate absolute affinity prediction or the physical validity of poses, and it remains unverified whether Boltzina can match Boltz-2 on these tasks.
In addition, rigid docking with AutoDock Vina cannot account for protein flexibility, and pocket selection requires prior knowledge.
Finally, applying the proposed method to ultra-large screening~\cite{Yu2023-wz,Correa-Verissimo2025-ka} exceeding one billion compounds still presents challenges.

Overall, the proposed method bridges the gap between Boltz-2's high accuracy and the speed of conventional docking, substantially improving cost-effectiveness in virtual screening and contributing to more efficient drug discovery.

\section*{Acknowledgements}
This study was partly supported by JSPS KAKENHI (JP23H04880, JP23H04887, JP24KJ1091), AMED BINDS (JP24ama121026), and JST FOREST (JPMJFR216J).

\bibliographystyle{unsrt}
\bibliography{main}

\end{document}